\documentclass[twoside]{article}
\usepackage[latin1]{inputenc}
\usepackage{t1enc}
\usepackage{a4}
\usepackage{tabularx}
\usepackage{epsf}
\usepackage{amssymb}

\def\bref{\vspace{4pt}\noindent\hangindent=10mm}

\begin{document}

\setcounter{figure}{0}
\setcounter{section}{0}
\setcounter{equation}{0}

\begin{center}
{\Large\bf
What cluster gas expulsion can tell us about \\
star formation, cluster environment \\ [0.15cm]
and galaxy evolution}\\[0.7cm]

Genevi\`eve Parmentier \footnote{Research Fellow of the {\it Alexander von Humboldt} Foundation}\\[0.17cm]
Argelander-Institut f\"{u}r Astronomie, Bonn Universit\"{a}t \\
Auf dem H\"{u}gel, 71 -- D-53121 Bonn -- Germany \\
gparm@astro.uni-bonn.de
\end{center}

\vspace{0.5cm}

\begin{abstract}
\noindent{\it
Violent relaxation -- the protocluster dynamical response to the expulsion of its leftover star 
forming gas -- is a short albeit crucial episode in the evolution of star clusters and star cluster
systems.  In this contribution, I survey how it influences the cluster age distribution, the cluster
mass function and the ratio between the cluster mass and the stellar mass.  I highlight the promising
potential that the study of this phase holds in terms of deciphering star cluster formation and galaxy evolution, and (some of) the issues which are to be dealt with before achieving this goal. 
}
\end{abstract}

\section{Introduction}
\label{sec:intro}
Although most stars in the local Universe form in star clusters, the stellar 
content of galaxies is dominated by field stars.  A thorough 
understanding of the lifecycle of star clusters, specifically, how and when they 
dissolve, is therefore of paramount importance to translate meaningfully their observed 
properties into the evolutionary history of their host galaxies. 

The evolution of star clusters is dominated by two main
phases: {\it (i) violent relaxation} followed by {\it (ii) secular evolution}.
Violent relaxation is the dynamical response of the protocluster to the expulsion
of its leftover star forming gas triggered by massive star activity (ionization overpressure,
stellar winds, Type II supernovae).  {\it If} it survives this first phase, the 
gas-free star-depleted cluster is back into virial equilibrium within 50\,Myr at most.  
It then enters secular evolution over which it steadily loses stars (evaporation), 
owing to the combined effects of mass loss through stellar evolution, internal 2-body 
relaxation and external tidal stripping, until its complete dissolution 
(see e.g. Meylan \& Heggie 1997, Kroupa, Aarseth \& Hurley 2001).

We now have a fairly mature and quantitative view of cluster
secular evolution, both empirically (Boutloukos \& Lamers 2003, Lamers et al.~2005) 
and theoretically (Baumgardt \& Makino 2003).  Its duration mostly depends on the cluster 
mass at the onset of secular evolution (i.e. at the end of violent relaxation) 
and on the cluster environment.  While the external tidal field induced by the smooth
host galaxy gravitational potential is often considered as the prime driver of the evolution 
of a population of star clusters, other -- local -- factors such as giant molecular clouds (GMC) 
also play a key-role in shortening cluster lifetimes
(Gieles et al.~2006).  The temporal decrease of the mass
of a cluster through secular evolution is conveniently described by, e.g., eq.~6
in Lamers et al.~(2005) which matches cluster $N$-body simulation outputs very well 
and where all cluster environmental effects (host galaxy tidal field, close passages 
with GMCs, spiral arm crossing, cluster orbit
eccentricity, ...) are encompassed by the parameter $t_4^{dis}$, an 
estimate of the dissolution time-scale of a cluster with a mass of $10^4\,{\rm M_{\odot}}$
at the onset of secular evolution. 

Modelling of cluster secular evolution has proved a powerful tool
to constrain, based on cluster present-day properties, the initial characteristics 
of star cluster systems.  That is, based on the observed age and mass distributions
of star clusters in a given (region of a) galaxy, we are able to disentangle the 
rates of formation and evolution of clusters which survived their violent relaxation 
(hereafter bound cluster; see e.g. Parmentier \& de Grijs 2008 for a thorough discussion).  
This leads to the history of the bound cluster formation rate (CFR) and to an estimate 
of the $t_4^{dis}$ parameter,
i.e. a measure of the bound cluster dissolution rate as driven by the cluster environment.  
It is worth keeping in mind, however, that the $t_4^{dis}$ parameter must have varied over
time since galaxies evolve, interact and merge, thereby modifying their mean density,
GMC abundance and structure.  Deriving a bound CFR
history based on a single estimate of $t_4^{dis}$ -- assumed constant over a Hubble-time --
thus remains fraught with uncertainties at old ages, when a galaxy may have been
very different to what it is today. 

The duration of secular evolution depends on the earlier violent relaxation
through cluster mass losses triggered by residual star forming gas expulsion.  
Due to gas expulsion, the young star cluster suddenly finds itself residing in a new
shallower gravitational potential (i.e., cluster stars are now moving too 
quickly for the gas-free cluster potential), which entails its expansion and 
the loss of a fraction of its stars ("infant weight-loss") or, even, its complete disruption
("infant mortality").  

While the first attempts of modelling cluster violent relaxation
dates back to the eighties, with the pioneering works of Hills (1980) and Lada, 
Margulis \& Dearborn (1984), it is recently only that we started to realize
the wide-ranging and far-ranging implications of this phase.  
Violent relaxation leaves an imprint on the structure of both young 
(Bastian \& Goodwin 2006) and old star clusters (Marks, Kroupa \& Baumgardt 2008), 
as well as on cluster system properties (e.g., their mass function and age 
distribution: Parmentier et al.~2008, Parmentier \& Fritze 2009).
Besides, violent relaxation is the key process which helps match two seemingly 
opposed paradigms: that most stars form in a clustered mode on the one hand, while the 
stellar content of galaxies is (by far) field-star-dominated on the other hand.  
It may also contribute to the formation of galactic structures such as galaxy thick 
discs (Kroupa 2002, 2005).

In spite of its shortness ($\lesssim 50$\,Myr), violent relaxation is thus a crucial phase of star cluster 
evolution which roots the physics of star cluster formation and evolution at an exciting 
border between star formation and galaxy evolution.  Deciphering and understanding the 
observational signatures of cluster gas expulsion and violent relaxation will tell us a 
lot about both processes, full stop.

\section{Ingredients of cluster gas expulsion modelling}
\label{sec:gr}

In the course of cluster violent relaxation, the instantaneous mass
of a star cluster obeys:
\begin{equation} 
m_{\rm cl}(t) = F_b(t, \varepsilon,\tau _{\rm GR}/\tau _{\rm cross}, r_h/r_t) \times \varepsilon \times m_{\rm c}\;.
\label{eq:m_inst}
\end{equation}
In this equation, $m_c$ is the mass of the cluster-progenitor gas core, $\varepsilon$ is the 
local star formation efficiency (SFE) and $F_b$ is the stellar mass fraction of the initially gas-embedded 
cluster which is still bound to the cluster (i.e. which still resides within the instantaneous 
cluster tidal radius).  

Formally, $\varepsilon$ is the {\it effective} SFE (eSFE), that is, the ratio of the stellar mass to the gas mass within the volume occupied by the stars at the moment gas expulsion starts (Verschueren 1990, Goodwin 2008).  This eSFE may be different from the {\it local/intrinsic} SFE, that is, the cluster-forming core gas mass fraction actually turned into stars.  This happens if the stellar system has evolved dynamically from the onset of star formation to gas expulsion, e.g. when cluster stars form subvirial, leading to a protocluster collapse and, thus, to an eSFE larger than the local/intrinsic SFE (Goodwin 1997).  In what follows, we assume that the stars have had sufficient time to come into virial equilibrium with the gas potential (i.e. gas expulsion occurs after a few crossing times), and so the eSFE very closely matches the local SFE.  

The bound fraction $F_b$ depends on the time $t$ elapsed since cluster gas expulsion, on the local SFE $\varepsilon$, on the gas removal time-scale $\tau _{\rm GR}/\tau _{\rm cross}$ (expressed in unit of a protocluster crossing-time) and on the impact  of the external tidal field, here described as the ratio  of the half-mass radius to the tidal radius of the gas-embedded (i.e. not yet expanded) cluster $r_h/r_t$.  The older the cluster and/or the lower the SFE and/or the quicker gas expulsion and/or the stronger the host galaxy tidal field, the lower the bound fraction $F_b$.
It is important to bear in mind that, in spite of its shortness, violent relaxation {\it is} affected by an external tidal field.  Evidence for this is provided by Scheepmaker et al.~'s ({\it subm.}) recent study of the M51 star cluster system.  They find that by an age of 10\,Myr, about 70\,\% of the M51 stellar content already resides in the field.  If the external tidal field were weak, the violent relaxation-driven scattering of stars out of clusters (i.e. beyond their tidal radius) is expected to start {\it from} an age of 10\,Myr (Parmentier \& Fritze 2009, their figs.~1 and 2).  M51 shows a strong tidal field, however, which, by virtue of the smaller cluster tidal radii, accelerates cluster dissolution within 10\,Myr following gas expulsion, in qualitative agreement with what is observed (see also section \ref{subsec:cmf} and Fig.~\ref{fig:tf_fb}).

The modelling of whole star cluster systems, in vast numbers so as to browse the parameter space extensively, has recently been rendered doable by the comprehensive $N$-body model grid of Baumgardt \& Kroupa (2007).  It provides the temporal evolution of the bound fraction $F_b(t, \varepsilon,\tau _{\rm GR}/\tau _{\rm cross}, r_h/r_t)$ over $1,000$ protocluster crossing-times following gas expulsion.
The $N$-body simulations of Baumgardt \& Kroupa (2007) were performed with $20,000$ equal-mass particles and assume that the Plummer mass density profile of the embedded cluster and of the cluster-forming core are alike, i.e. the SFE is independent of the location within the cluster-progenitor gas core (see Adams 2000 for an iterative analytical approach to obtain the bound fraction when the local SFE increases towards the cluster-forming core centre).  Such an extensive model grid allows for significant deeper insights into the early evolution of star cluster systems, whose integrated properties (age/mass/radius distribution functions) can now be recovered by interpolating the $N$-body model grid, rather than $N$-body modelling each cluster of each fiducial cluster system.  Future improvements on these results include systematic investigations of the effects induced by a stellar initial mass function, the presence of primordial binaries, of cluster primordial mass segregation and of cluster density substructures in general (see Marks, Kroupa \& Baumgardt 2008 for insights into some of these aspects). 

\begin{figure}[t]
\begin{center}
\epsfxsize=10cm
\epsfbox{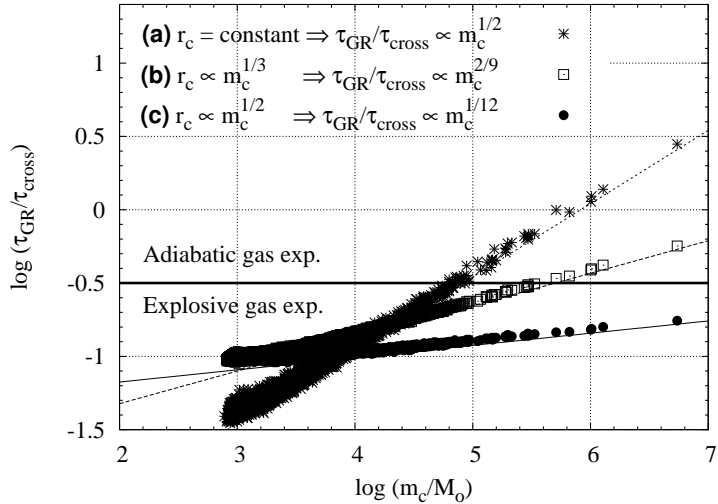}
\caption{Relations between gas expulsion time-scale $\tau _{GR}$ in unit of 
one protocluster crossing time $\tau _{cross}$ and cluster-forming core mass $m_c$, 
for 3 core mass-radius relations $m_c-r_c$.  Figure 3 in Parmentier \& Fritze (2009).}
\label{fig:tau_GR}
\end{center} 
\end{figure}

Albeit not conspicuous in Eq.~1, the instantaneous cluster bound fraction $F_b$ depends on the mass $m_c$ and radius $r_c$ of the cluster-forming core, through both the gas removal time-scale and the tidal field.  For a given cluster environment, the more compact the cluster, the smaller the $r_h/r_t$ ratio and the likelihood that the expanding cluster will overflow its tidal radius significantly.  Besides, a greater compactness also implies a larger cluster-forming core mass density, thereby slowing down gas expulsion and limiting in turn cluster spatial expansion and infant weight-loss (Baumgardt \& Kroupa 2007).

Based on how long a supershell takes to sweep the leftover core gas out of the protocluster, Parmentier et al.~(2008) derive an expression for the gas expulsion time-scale $\tau _{\rm GR}/\tau _{\rm cross}$ (their eq.~6; see also Baumgardt, Kroupa \& Parmentier 2008 for an alternative estimate of the gas expulsion time-scale based on the amount of mechanical energy injected into the interstellar medium by protocluster massive stars).  Figure \ref{fig:tau_GR} displays how $\tau _{\rm GR}/\tau _{\rm cross}$ scales with the core mass $m_c$ for three different core mass-radius relations $m_c-r_c$:
constant core radius $r_c$ (case {\it a}: $r_c=0.7$\,pc), constant core volumic density (case {\it b}: $r_c=0.026 \times (m_c/1M_{\odot})^{1/3}$\,pc) and constant core surface density (case {\it c}: $r_c=0.008 \times (m_c/1M_{\odot})^{1/2}$\,pc).  Normalizations of these three relations are based on fig.~1 of Tan (2007) which shows the radius, mass and surface density of a sample of {\it Galactic} protoclusters (see section 2 of Parmentier \& Fritze 2009 for details).  In case {\it a}, the core mass sequence equates with a core mass density sequence which causes higher mass cores to experience slower gas expulsion (in protocluster crossing-time units) and, therefore, to retain a larger bound fraction $F_b$.  In contrast, in case of a constant core surface density (case {\it c}), the gas expulsion time-scale and the bound fraction are almost insensitive to core mass.  The nature of the core mass-radius relation is therefore tightly related to whether cluster infant mortality is mass-dependent or not (see also section \ref{subsec:cmf}).

\section{Gas expulsion and cluster system properties}
\label{sec:gr_cons}
\subsection{Cluster age distributions}
\label{subsec:im}

From Eq.~\ref{eq:m_inst}, it follows that cluster infant mortality rates depend on the 
nature of cluster-forming cores -- their mass-radius relation --, star formation --~its 
efficiency -- and cluster environment -- the host galaxy.  The survey of a substantial 
number of young star cluster systems so as to analyse the relative influences from cluster
formation and cluster environment on infant mortality rates constitutes a task of paramount 
importance.  

In Parmentier \& Fritze (2009), we set off to investigate the impact of the local SFE and of
the core mass-radius relation on the cluster age distribution in the case of a weak tidal field.
Specifically, we trace the temporal evolution of the mass in star clusters, which 
has the advantage of quantifying the impact of {\it both} infant mortality and infant weight-loss,
over their first 100\,Myr (their figs.~1 and 2).  We adopt different Gaussian distribution 
functions, $G(\bar\varepsilon, \sigma _{\epsilon})$, 
of the local SFE and the core mass-radius relations quoted above.
We show that the ratio between the total mass in stars bound to the clusters (i.e. located within cluster instantaneous tidal radii) and the total mass in stars formed in gas-embedded clusters, integrated over the age range 1-100\,Myr, is a remarkably sensitive tracer of the mean SFE, $\bar\varepsilon$.  This is expected since, for an individual protocluster, the bound fraction of stars at the end of violent relaxation is a sharply increasing function of the local SFE (see fig.~1 in Parmentier \& Gilmore 2007 for the case of a weak tidal field).  Besides, this cluster-to-star mass ratio shows only a weak dependence to fine model details such as the slope of the core mass-radius relation or the standard deviation $\sigma _{\epsilon}$ of the assumed Gaussian SFE distribution (Table 1 in Parmentier \& Fritze 2009).  Our result suggests that measurements of the flux ratio from star clusters relative to field stars constitute a promising way of probing the local SFE in active star forming environments, without resorting to gas mass estimates.  Prerequisites include that the age range of the star cluster system includes the post-violent-relaxation age range, i.e. [50,100]\,Myr for a weak tidal field. 
 
{\it Figures quoted in Table 1 of Parmentier \& Fritze (2009) should not be blindly applied to observational cases, however}, as they still relie on a number of simplifying assumptions such as the constancy of the gas-embedded CFR -- equivalent to the SFR if most stars form in gas-embedded clusters -- and a weak tidal field
(equivalent to that in the Milky Way halo at a Galactocentric distance of about 40\,kpc).  While generalized versions of our Monte-Carlo-based model can be developed straightforwardly, to estimate the strength of a tidal field and the recent (i.e. over the last 100\,Myr) time-varying SFR of an interacting galaxy may prove more challenging.  Besides, any observed cluster-to-star mass ratio is to be corrected for observational biases (e.g. detection limit and crowding effects).  Finally, it must be kept in mind that the cluster-to-star mass ratio robustness quoted above stems from the inclusion of clusters not yet significantly affected by violent relaxation (i.e. ages younger than 10\,Myr).  Should the cluster-to-star mass ratio be integrated over the age range 50-100\,Myr, so as to encompass violent relaxation survivors only, it would appear much more sensitive to the core mass-radius relation, the slope of which remains debated.  This is illustrated in Table 1 of this contribution.  For a mean local SFE of 40\,\%, the uncertainty affecting the core mass-radius relation alone leads to a relative uncertainty of about 50\,\% for the bound-cluster-to-star mass ratio.  Following the recent finding by Bastian (2008) that galaxies form, on the average, 8 per cent of their stars in bound clusters regardless of their SFR, we raise the hypothesis that star formation {\it in the present-day Universe} is characterized by a near-universal mean SFE.  I am currently furthering this hypothesis through the modelling of the relation between the SFR of galaxies and the absolute visual magnitude of their brightest young cluster.  

\begin{table}
\begin{center}
\caption{Ratio between the total mass in bound clusters and the total mass in stars integrated over the age range 50-100\,Myr, for various Gaussian distribution functions $G(\bar\varepsilon,\sigma _{\varepsilon})$ of the local SFE and core mass-radius relations.  A weak tidal field and a constant gas-embedded CFR are assumed.  Compare this table to table 1 in Parmentier \& Fritze (2009), where integrations of cluster and stellar masses are performed over the age range 1 to 100\,Myr}
\smallskip
\smallskip
\begin{tabular}{c c c c c c c c c} 
\hline 
                  &    &  \multicolumn{3}{c}{$\bar \varepsilon = 0.25$} & & \multicolumn{3}{c}{$\bar \varepsilon = 0.40$}   \\ 
\cline{3-5} \cline{7-9} 
$\sigma _\varepsilon$ & ~~~ & ~$0.01$~ & ~$0.04$~ & ~$0.07$~ & ~~~ & ~$0.01$~ & ~$0.04$~ & ~$0.07$~  \\ 
\hline
$ r_c = constant $      &     &  $0.03$  &  $0.04$  &  $0.09$  &     &  $0.35$  &   $0.35$ &  $0.38$   \\

$ r_c \propto m_c^{1/3} $ & ~~~ & ~$0.00$~ & ~$0.00$~ & ~$0.02$~ & ~~~ & ~$0.27$~ & ~$0.26$~ & ~$0.30$~  \\ 

$ r_c \propto m_c^{1/2} $  & ~~~ & ~$0.00$~ & ~$0.00$~ & ~$0.01$~ & ~~~ & ~$0.21$~ & ~$0.23$~ & ~$0.29$~  \\
 \hline

\end{tabular}
\end{center}
\end{table}  

Another issue of potential concern is that our model assumes that {\it all} stars form in gas-embedded clusters.  It remains ill-known whether a genuine mode of distributed star formation exists or not.
Using the best presently available kinematic data on O-stars and young open clusters, Schilbach \& R\"{o}ser (2008) back-trace the orbits of 93 O-type field stars and young open clusters, and find 73 O-stars to originate from 48 open clusters younger than 30 Myr.  This is a lower limit since the parent clusters of some field O-stars may be dissolved by now (see also Gvaramadze \& Bomans 2008 for the origin of some field O-stars in clusters as evidenced by bow shocks arising from their supersonic motion through the ambient interstellar medium).  Schilbach \& R\"{o}ser 's (2008) result thus shows that most field O-stars form in clusters rather than in isolation.  That star clusters may be surrounded by a more diffuse mode of star formation is to be expected however, since cluster winds may trigger the formation of new stars in the compressed shell of gas they have swept away (Parmentier 2004), but the limited mass of gas in each fragment formed along the shell periphery may prevent the formation of massive stars there.  This diffuse mode may, however, constitute small Taurus-Auriga-like groups or "clusters" of T Tauri stars.  According to Lada \& Lada (2003), in the Galactic disc, about 70-90\,\% of stars form in open clusters.  In Parmentier \& Fritze (2009), we demonstrate that a diffuse mode of star formation, which accounts for 10-30\,\% of all star formation, does not affect significantly the local SFE estimate one would retrieve from the observed cluster-to-star mass ratio, in essence because of the sharp increase of the bound fraction with the local SFE.

\subsection{Cluster mass functions}
\label{subsec:cmf}

\begin{figure}[t]
\begin{center}
\epsfxsize=10cm
\epsfbox{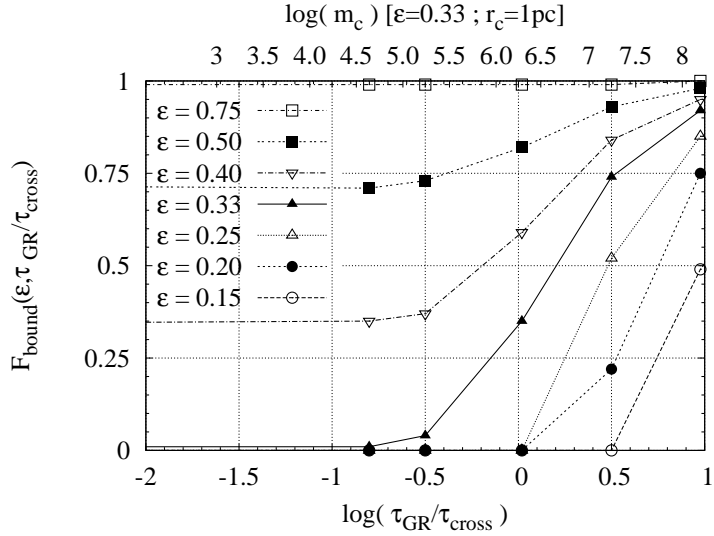}
\caption{Mass fraction $F_{\rm bound}$ of stars remaining bound to the
cluster {\it at the end of violent relaxation} in case of a weak tidal field
(i.e. at a cluster age of $\simeq 50Myr$).   $F_{\rm bound}$ varies with the 
local SFE $\varepsilon$ and with the ratio between  the gas removal time-scale 
${\tau _{GR}}$ and the core crossing-time  ${\tau _{cross}}$.  
The top and bottom $x$-axes match is provided by eq.~6 in Parmentier et al.~(2008).
Figure 1 in Parmentier et al.~(2008)} 
\label{fig:fb_tau_sfe}
\end{center} 
\end{figure}

That the cluster mass function (CMF) may retain an imprint of gas expulsion physics was first suggested by Kroupa \& Boily (2002).
Their model builds explicitely on the hypothesis that cluster-forming cores have the same radius, irrespective of their mass.  Assuming a unique local SFE of about 30\,\%, they show that a power-law core mass function evolves into a bell-shaped CMF, when considering clusters whose violent relaxation is over. 

\begin{figure}[t]
\begin{center}
\epsfxsize=10cm
\epsfbox{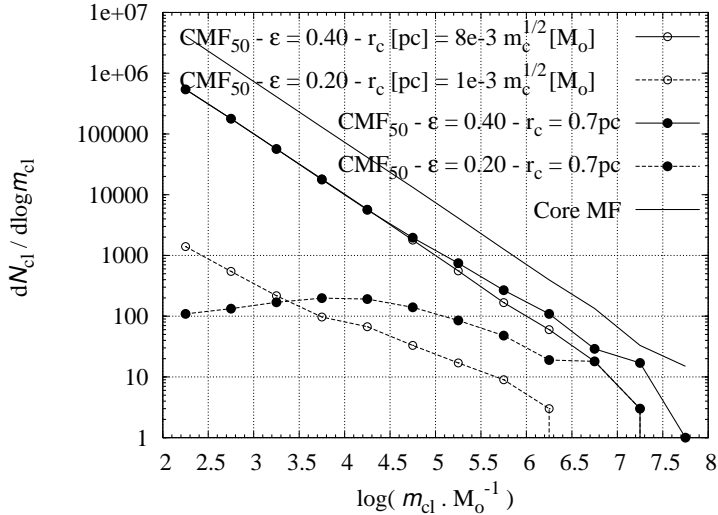}
\caption{Impact of mean SFE and of core mass-radius relation on the shape of the CMF when violent relaxation is over, i.e.~at an age of 50\,Myr ("CMF$_{50}$") for a weak tidal field as assumed here.  The standard deviation of the SFE Gaussian distribution is $\sigma _{\varepsilon} = 0.03$.  Note that the normalization of the core mass-radius relation matters too.  Adapted from Parmentier et al.~(2008)}
\label{fig:cmf}
\end{center} 
\end{figure}

Building on the $N$-body model grid of Baumgardt \& Kroupa (2007), Parmentier et al.~(2008) revisit the issue, exploring the parameter space of the SFE distribution function and of the core mass-radius relation.  The essence of our results is highlighted in Fig.~\ref{fig:cmf}.  \\
{\bf (a)} Constant (i.e.~mass-independent) core radius: assuming that cluster-forming core mass functions are featureless power-laws, bell-shaped CMFs arise when the mean local SFE is smaller than 25\,\%.  In contrast, SFE $\gtrsim$ 40\,\% preserve the power-law shape while intermediate values (say, SFE $\simeq 30$\,\%) result in CMFs flatter than the core mass function (see also Baumgardt et al.~2008).  The turnover or the flattening of the CMF compared to that of cores results from the deeper gravitational potential of larger mass cores causing them to experience slower gas expulsion (in units of the protocluster crossing time) and, thus, to retain a higher bound fraction $F_{bound}$ of stars at the end of violent relaxation (see Fig.~\ref{fig:fb_tau_sfe}).  When the local SFE is large (say, 40\,\% or larger), however, the dependence of $F_{bound}$ on the core mass $m_c$ is not sensitive enough to alter the shape of the relaxed CMF.  \\
{\bf (b)} The core radius $r_c$ is an increasing function of $m_c$ ($r_c \propto m_c^{1/2}$): the gas removal time-scale dependence on $m_c$ weakens (see eq.~6 in Parmentier et al.~2008) and so does the dependence of the bound fraction on the core mass.  Consequently, the power-law core mass function evolves into a power-law CMF of similar slope.  One may be puzzled by the presence in Fig.~\ref{fig:cmf} of a CMF when $\varepsilon = 20$\,\% (dashed line with open circles), since Table 1 quotes a null bound-cluster-to-star mass ratio in such a case.  The difference between Table 1 and Fig.~\ref{fig:cmf} resides in the normalization of the core mass-radius relation: $r_c {\rm [pc]} = 0.008 (m_c/{\rm M_{\odot}})^{1/2}$ and $r_c {\rm [pc]} = 0.001 (m_c/{\rm M_{\odot}})^{1/2}$, respectively.  That is, the core density is higher in the ($\varepsilon = 20$\,\%, $r_c \propto m_c^{1/2}$) case of Fig.~\ref{fig:cmf} than in Table 1, which slows down gas expulsion and increases the survivability of gas-embedded clusters.  This example demonstrates that, not only does the slope of the core mass-radius relation matter, its normalization also influences the outcome of violent relaxation.

Observed young CMFs have so far been reported to be power-laws, although CMF slopes vary (from $-1.8$ to $-2.4$ ) from one study to another (see also Larsen 2009 for evidence for a Schecter mass function with an environmentally-dependent cut-off mass).  Based on modelling results described above, it is interesting to note that, {\it if core radii are constant}, the absence of substructures in -- so far -- observed CMFs may tell us something about the local SFE, a point discussed in the next section.

A power-law young CMF is at variance with the observed old globular CMF, since secular evolution, which leads to the preferential removal of low-mass clusters, may prove unable to evolve a power-law CMF in a Gaussian CMF with a turnover at a cluster mass of $2 \times 10^5\,{\rm M}_{\odot}$ within a Hubble-time (Parmentier \& Gilmore 2005, Vesperini et al.~2003; but see Jordan et al.~2007 for an alternative point of view).  ``Special'' conditions, which may be specific to the protogalactic era, have therefore been invoked such as a characteristic protoglobular cloud mass of $\simeq 10^6\,{\rm M}_{\odot}$, subsequently turned into the universal globular CMF by violent relaxation (Parmentier \& Gilmore 2007), or a SFE of 25\,\% which flattens the relaxed CMF (i.e. violent relaxation alone already removes a substantial relative fraction of low-mass clusters; Baumgardt et al.~2008).  Deep (say, down to a few $10^3\,{\rm M}_{\odot}$) intermediate-age (1--2\,Gyr) CMFs should help settle this puzzling and enduring issue.

\begin{figure}[t]
\begin{center}
\epsfxsize=10cm
\epsfbox{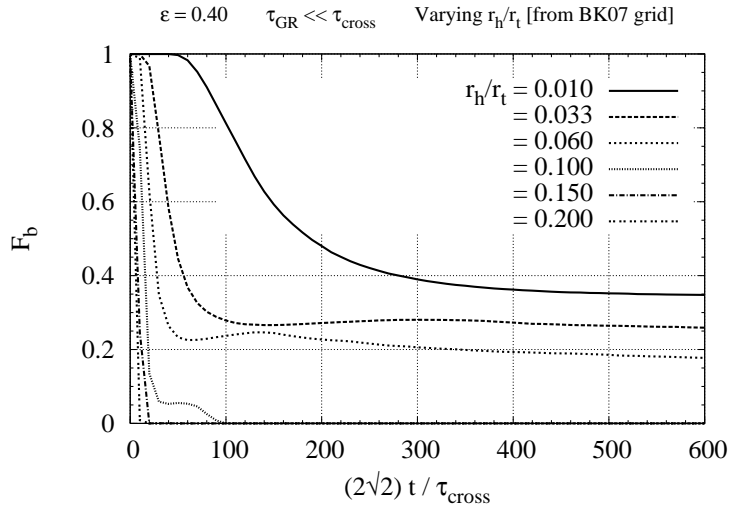}
\caption{Temporal evolution of the bound fraction of stars in case of 40\,\% SFE and instantaneous gas expulsion.  The different lines correspond to diffferent half-mass radius to tidal radius ratios of the gas-embedded cluster }
\label{fig:tf_fb}
\end{center} 
\end{figure}

Finally, it is worth keeping in mind that model and observed CMFs must be compared meaningfully, {\it i.e. at the same evolutionary stage}.  If core radii are constant, that the CMF shape bears clues about the local SFE is true only for clusters which have completed their violent relaxation.  It is therefore of paramount importance to build observed CMFs at that evolutionary stage.  If the observed CMF encompasses all clusters, including the very young ones still mildly affected by infant weight-loss/mortality, the power-law gas-embedded CMF (i.e. the distribution of clusters with masses $\varepsilon \times m_c$, which mirrors the core mass function if the SFE is core-mass independent) "contaminates" the relaxed CMF, which may conceal substructures in the latter, especially in star cluster systems with high infant mortality rates.  Estimates of the age of relaxed clusters are often quoted in the range 30--50\,Myr.  This is appropriate only for a weak tidal field, however.  Because the tidal radius of star clusters is smaller in stronger tidal fields (for a given cluster mass), unbound stars reach the cluster tidal boundary within a shorter time span.  To define the end of cluster violent relaxation is not that straightforward as it depends on the tidal field strength and is, therefore, environment-dependent.  This point is illustrated in Fig.~\ref{fig:tf_fb} which shows the temporal evolution of the instantaneous bound fraction of stars for different ratios $r_h/r_t$ between the half-mass radius and the tidal radius of gas-embedded clusters when $\varepsilon=40$\,\% and gas expulsion proceeds explosively (i.e. $\tau _{\rm GR} << \tau _{\rm cross}$).  All curves are extracted from the $N$-body model grid of Baumgardt \& Kroupa (2007).  The complexity of estimating cluster violent relaxation duration is now clearly highlighted as it depends on the cluster mass and the external tidal field through $r_t$, and on the assumed core mass-radius relation through $r_h$.  Besides, note that the choice of the core mass-radius relation also affects the conversion of $N$-body time units into physical time units.  Simulations encompassing various tidal field strengths will be presented in forthcoming papers.

\section{Importance of the core mass-radius relation}
\label{sec:mcrc}

\subsection{An interesting conundrum}
\label{susec:conun}

The exact nature of the core mass-radius relation has kept going through an enduring crisis.
Virialised gas cores are expected to show a strong mass-radius relation scaling as 
$r_c \propto m_c^{1/2}$ (e.g. Harris \& Pudritz 1994).  The mass-radius relation of young clusters is therefore expected to display the same trend, albeit with much scatter since age and local SFE cluster-to-cluster variations give rise to different degrees of cluster expansion.  An interesting consequence of this assumption is that the gas removal time-scale $\tau _{GR}/\tau _{cross}$, thus the bound fraction $F_{b}$, are practically core-mass independent (case {\it c} in Fig.~\ref{fig:tau_GR}), leading to mass-independent infant mortality rates, as observed thus far (e.g.~Bastian et al.~2005).  Yet, a sound observational fact is that star clusters -- both young clusters and old globulars -- show no clear-cut mass-radius relation, i.e. $r_{cluster} \propto m_{cluster}^{0-0.1}$ (e.g. Larsen 2004).  Kroupa (2005) therefore proposed that, contrary to theoretical expectations, cluster-forming-cores are characterized by a constant radius over their entire mass range.  This accounts naturally for the observed cluster radius constancy (albeit with much scatter), but may seem at variance with cluster-mass-independent infant mortality rates inferred so far since high-mass gas-embedded clusters have then a higher likelihood of survival.

In the frame of the $r_c \propto m_c^{1/2}$ hypothesis, Ashman \& Zepf (2001) proposed that lower mass cores are characterized by smaller local SFEs than high mass ones, causing them to achieve, following gas expulsion, greater spatial expansion than high mass cores.  As a result, the initial core mass-radius relation is wiped out and the CMF gets flatter over the course of violent relaxation
\footnote{Although the flattening effect on the CMF is similar to what has been found by Parmentier et al.~(2008), its origin differs.  In Ashman \& Zepf (2001), it is due to higher mass-cores having higher local SFEs, while in Parmentier et al.~(2008), it results from higher mass cores experiencing slower gas-expulsion.}.  Although their flattened CMF remains reasonably compatible with observed CMF slopes, the amount of CMF flattening they obtained is likely underestimated as their model does not account for cluster infant weight-loss (i.e. $F_{bound} =1$ in their model).  Whether the local SFE actually increases with or is independent of core mass is observationally poorly-known since local SFEs have been estimated for a handful only of nearby Galactic gas-embedded clusters (Lada \& Lada 2003).
As for the $r_c=constant$ hypothesis, the observed mass-independent infant mortality rate of star clusters can be accommodated provided a high enough local SFE, say, $\gtrsim$35\,\%.  In this case, the bound fraction $F_{bound}$ depends weakly only on the core mass $m_c$ (Fig.~\ref{fig:fb_tau_sfe}) and a power-law core mass function evolves into a relaxed CMF of similar slope (Fig.~\ref{fig:cmf}).

Studying the cluster mass-radius relation as a function of age is probably an instructive exercice to perform.  Note that this task brings us back to defining the duration of cluster violent relaxation (see section \ref{subsec:cmf}).  The advent of ALMA will undoubtedly settle the core mass-radius relation issue.  Till then, the observed properties of young clusters coupled to violent relaxation modelling remain our best proxy to the properties of dense cluster-forming cores of extragalactic GMCs.

\subsection{Cluster observables as probes to galaxy evolution and star formation}
\label{susec:gal}

\begin{figure}[t]
\begin{center}
\epsfxsize=10cm
\epsfbox{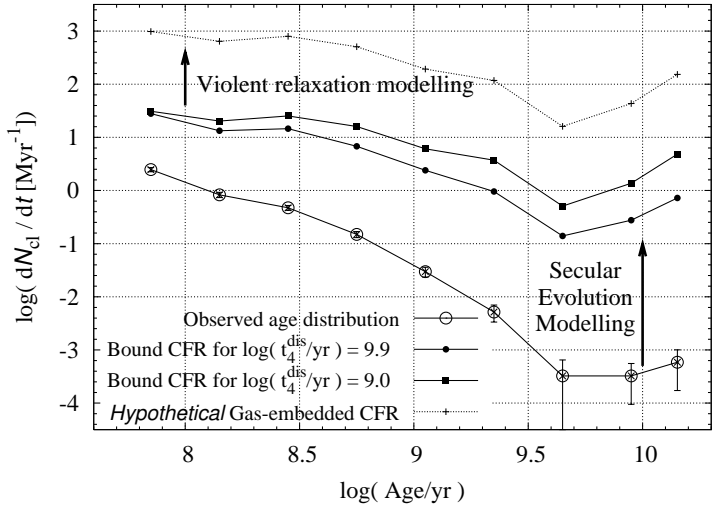}
\caption{From the observed age distribution of star clusters to the SFR of a galaxy.  Modelling of secular evolution allows to recover the bound CFR.  Violent relaxation modelling may enable us to infer one step further the history of star cluster formation and to derive the formation rate of gas-embedded clusters.  Observed age-distribution and bound CFR histories are those of the Large Magellanic Cloud.  Adapted from Parmentier \& de Grijs 2008}
\label{fig:lmc}
\end{center} 
\end{figure}

As was already discussed in Section \ref{subsec:cmf}, the shape of the CMF at the end of violent relaxation constitutes a valuable indicator of the mean local SFE only if core radii are mass-independent.  
Another cluster observable affected by the slope of the core mass-radius relation is the bound-cluster-to-star mass ratio (section \ref{subsec:im} and Table 1), that is, the ratio betwen the total mass in clusters at the end of violent relaxation and the total mass in stars formed in the same star formation episode.  If the core mass-radius relation were known, one could, for a galaxy whose field stars can be resolved and age-dated, infer the history of the bound-cluster-to-star mass ratio and, in turn, information about cluster violent relaxation over galaxy evolution, such as the history of the mean local SFE.  
Note that the history of the host galaxy tidal field should be known (or assumed) too.  Conversely, the bound-cluster-to-star mass ratio has the potential of tracing the SFR history of galaxies (Maschberger \& Kroupa 2007).  As quoted in the introduction, secular evolution modelling -- coupled with correction of fading effects -- allows us to lookback in time and to infer, from the observed cluster age distribution, the age distribution of clusters at the onset of secular evolution, i.e. the age distribution of clusters which emerged bound from their violent relaxation.  A firm grasp on violent relaxation would allow us to go one step further and to infer, from the recovered bound cluster age distribution, the formation rate history of gas-embedded clusters, that is, to correct the bound CFR for the disruption of some of the gas-embedded clusters and infant weight-loss of the survivors.  Note that if most stars form in clusters, the gas-embedded CFR
practically equates with the overall SFR.  This concept is illustrated in Fig.~\ref{fig:lmc}, where the observed cluster age distribution and the bound CFR history of the Large Magellanic Cloud are taken from Parmentier \& de Grijs (2008).  A related issue is the existence of gaps in the observed age distribution of star clusters, e.g. in the Large Magellanic Cloud over the age range 3-10\,Gyr.  Is it due to a low local SFE preventing any gas-embedded clusters from surviving their violent relaxation?  Or does it arise from a low SFR leading to the formation of low-mass clusters only (Weidner, Kroupa \& Larsen 2004) which have secularly dissolved by now~?

\section{Conclusions}

Cluster gas expulsion and violent relaxation are complex processes of which we have only started unravelling the various facets.  They are influenced by star formation (through the local SFE), the nature of cluster-forming cores (through the core mass-radius relation, both its slope {\it and} its normalization) and the environment, via the external tidal field.  Besides, the external pressure must influence the degree of compression of cluster-forming cores and, therefore, the gas expulsion time-scale.  The outcome of violent relaxation is a substantial fraction of stars, initially contained in gas-embedded clusters, scattered into the field.  Violent relaxation is therefore the link between star formation and the evolution of galaxies, most notably their star formation histories.  Significant leaps forward have been achieved over the past few years.  Yet, many questions remain unanswered, especially in terms of protocluster initial conditions.  Do star clusters form in virial equilibrium with their gas reservoir core~? 
Is there a relation between core mass and local SFE, or environment and local SFE~?  What is the slope of the  
cluster-forming core mass-radius relation~?  In this respect, it should be noted that the relation of relevance here is that of gas-embedded clusters {\it at the onset of gas expulsion.}  Whether a gas-embedded cluster dynamically evolves from star formation to gas expulsion and whether this affects the slope of the mass-radius relation are questions beyond the scope of the present discussion.

How quickly gas is expelled requires much deeper attention too since, in the highly dense protocluster environment, a fraction of the stellar wind/SNII energy is likely to be lost through stellar wind/shock wave collisions.  

Model testing requires comparison with observations.  In view of that, modelling of physical processes {\it and} observational biases is required.  For a cluster experiencing violent relaxation, how well can we distinguish the bound core of stars from the expanding halo of unbound ones~?  Is the instantaneous bound mass of a cluster as given by Eq.~\ref{eq:m_inst} a reliable match to a young cluster luminous mass estimate as obtained by an observer~?  Which fraction of the expanding cluster outskirts remains concealed by the background of its host galaxy ?  Much remains to be explored and explained, which will be achieved only via a tight collaboration between observers and modellers.  

\vspace*{-1mm}
\subsection*{Acknowledgements}
It is a great pleasure to thank my collaborators who contributed to part of the work described
above: P.~Kroupa, U.~Fritze, S.~Goodwin, H.~Baumgardt and R.~de Grijs.
I am grateful to the organizers -- E.~Pauzen, R.~Spurzem and P.~Kroupa -- for having invited me to give a highlight talk at this much enjoyable conference.  S.~Larsen and P.~Kroupa are acknowledged for useful comments about an earlier version of this manuscript.
This work has been supported by a Humboldt Fellowship.

\subsection*{References}

{\small

\bref
Adams, F.\,C. 2000, ApJ 542, 964

\bref
Ashman, K.\,M., Zepf, S.\, E. 2001, AJ 122, 1888

\bref
Bastian N., et al.~2005, A\&A 431, 905

\bref
Bastian, N., Goodwin, S.\,P. 2006, MNRAS: Letters 369, 9 

\bref
Bastian N. 2008, MNRAS 390, 759

\bref
Baumgardt, H., Makino, J. 2003, MNRAS 340, 227

\bref
Baumgardt, H., Kroupa, P. 2007, MNRAS 380, 1589 [BK07]

\bref
Baumgardt, H., Kroupa, P., Parmentier, G. 2008, MNRAS , 384, 1231

\bref
Boutloukos, S. G., Lamers, H. J. G. L. M. 2003, MNRAS 338, 717

\bref
Gieles, M., Portegies Zwart, S.\,F., Baumgardt, H. et al.~2006, MNRAS 371, 793

\bref
Goodwin, S. P. 1997, MNRAS, 284, 785

\bref
Goodwin S.\,P. 2008, in Young Massive Star Clusters: Initial Conditions and Environments,
ed. E.~Perez et al.~(Berlin: Springer), in press

\bref
Gvaramadze, V.\,V., Bomans D.\,J. 2008, A\&A, in press (arXiv:0809.0650v1)

\bref
Harris, W.\,E., Pudritz R.\,E. 1994, ApJ 429, 177

\bref
Hills, J.\,G. 1980, ApJ 225, 986

\bref
Jordan et al.~2007, ApJS 171, 101

\bref
Kroupa, P., Aarseth, S., Hurley, J. 2001, MNRAS 321, 699

\bref
Kroupa, P. 2002, MNRAS 330, 707

\bref
Kroupa, P., Boily, C.M. 2002, MNRAS, 336, 1188

\bref
Kroupa, P. 2005, In: Proceedings of "The Three-Dimensional Universe with Gaia" 
(ESA SP-576) C. Turon, K.S. O'Flaherty, M.A.C. Perryman (eds), p.629

\bref
Lada, C.\,J., Margulis, M., Dearborn, D. 1984, ApJ 285, 141

\bref
Lada, C.\,J., Lada, E.\,A. 2003,  ARA\&A 41, 57

\bref
Lamers, H.\,J.\,G.\,L.\,M., Gieles, M., Bastian, N.~et al.~2005, A\&A 441, 117

\bref
Larsen, S.\,S. 2002, AJ 124, 1393

\bref
Larsen, S., 2009, A\&A, in press (arXiv:0812.1400)

\bref
Larsen, S.\,S. 2004, A\&A 416, 537

\bref
Marks, M., Kroupa, P., Baumgardt, H. 2008, MNRAS 386, 2047

\bref
Maschberger, T., Kroupa, P. 2007, MNRAS, 379, 34

\bref
Meylan, G., Heggie, D. 1997, A\&ARv 8, 1

\bref
Parmentier, G. 2004, MNRAS 351, 585

\bref
Parmentier, G., Gilmore G. 2005, MNRAS 363, 326

\bref
Parmentier, G., Gilmore G. 2007, MNRAS 377, 352

\bref
Parmentier, G., de Grijs, R.. 2008, MNRAS 383, 1103

\bref 
Parmentier, G., Goodwin, S. P., Kroupa, P., Baumgardt, H. 2008, ApJ 678, 347

\bref
Parmentier, G., Fritze, U. 2009, ApJ 690, 1112 (arXiv0809.2416)

\bref
Scheepmaker R.\,A., Lamers, H.\,J.\,G.\,L.\,M., Larsen, S.\,S., Anders, P., 
subm.~to A\&A

\bref
Schilbach, E., R\"{o}ser, S. 2008, A\&A 489, 105

\bref
Tan, J.\,C. 2007, in: Triggered Star Formation in a Turbulent ISM, B.G. Elmegreen and J. Palous ed. -
Proc.~of IAU Symposium 237. CUP 2007, p258

\bref
Verschueren, W. 1990, A\&A 234, 156

\bref 
Vesperini, E., Zepf, S.\,E., Kundu A., Ashman, K.\,M. 2003, ApJ 593, 760

\bref
Weidner, C., Kroupa, P., Larsen, S.\,S. 2004, MNRAS 350, 1503


}

\vfill

\end{document}